\documentstyle[aps,twocolumn,psfig,epsf]{revtex}
\newcommand{\lan}{\langle}
\newcommand{\ran}{\rangle}
\begin{document}

\author{Mario Nicodemi}
\address{
\vspace{0.2cm}
Imperial College of Science, Technology and Medicine, 
 Department of Mathematics \\
 Huxley Building, 180 Queen's Gate, London, SW7 2BZ, U.K. 
$~~~$ nicodemi@ic.ac.uk \\
}

\title{Domains growth and packing properties in driven granular media 
subject to gravity} 

\maketitle
\date{\today}
\begin{abstract}

We study the dynamical properties of recently introduced frustrated 
lattice gas models (IFLG and Tetris) for granular media under gentle shaking. 
We consider both the case where grains have inter-grain surface 
interactions and the case where they have not, corresponding, for instance, 
to the presence or absence of moisture in the packs. 
To characterise the grains packing structure, 
we discuss the properties of density distribution. In particular, we consider 
the phenomenon of grains domains formation under compaction. 
New results amenable of experimental check are discussed 
along with some important 
differences between the dynamics of the present models. 
\end{abstract}

\section{Introduction}

Non-thermal disordered systems as granular media show 
a variety of dynamical phenomena with interesting properties 
\cite{JNBHM,Kadanoff}. 
For instance, important physical features were recently 
discovered in shaking experiments, where the dynamics of grains is driven, 
under the effects of gravity, by sequences of shakes (``taps'') \cite{Knight}. 
A very common phenomenon recorded in those experiments is compaction: 
when a box filled with loose 
packed sand is shaken at low amplitudes, $\Gamma_{ex}$
($\Gamma_{ex}=a/g$, where $a$ is the shake peak 
acceleration and $g$ the gravity acceleration constant), 
the bulk density of the system increases with the number of shakes. 
Interestingly, the Chicago group has experimentally shown that compaction 
is logarithmically slow \cite{Knight}. 
Several models have been proposed to describe these kind of dynamical 
behaviours
\cite{Mehta-Barker-Duke,CH,NCH,Ben-Naim,degennes,Head,Caglioti}, but 
a detailed investigations of its microscopic nature is still incomplete. 

In the present paper we discuss some aspects of such a problem by 
focusing on the details of the packing structure. 
Our analysis is developed in the context of 
recently introduced lattice gas models 
for granular media\cite{NCH,Caglioti,NC_aging}: the IFLG and Tetris. 
These models are very simple and thus very schematic, but they 
interestingly describe, in a single framework, many  
dynamical properties of granular media ranging from 
logarithmic compaction \cite{NCH}, ``irreversible-reversible" cycles 
and aging \cite{NC_aging}, anomalous dynamical responses 
\cite{N_fdt}, to segregation, avalanches effects 
and several others\cite{NC_rev}. 
As a generalisation of the q-Model \cite{Cop}, 
they have been also studied \cite{N_forces} 
to describe ``scalar force'' patterns. 

To understand the grain packing structure, we  
recorded the bulk density distribution in the present models, 
which in some cases turns out to be definitely non Gaussian. 
We propose a variant of our models to be able to describe also 
grain-grain surface interactions, which are to be considered for instance 
in cases of charged grains or non dry media. 
Interestingly, we find that the dynamics of compaction is not qualitatively 
affected by these details of grains interactions, at least up to when 
the coupling strength is of the order of the effective temperature 
induced by shaking. 
We have also studied some microscopic aspects of density compaction 
concerning the formation of ``domains of grains'' during the dynamics, 
which we describe in some details. 
All these results shed some light on the origin and the character 
of the slow dynamics in vibrated granular media. 
An experimental check of our findings, 
which in many cases is still missing, would be important to settle 
our theoretical understanding of granular media. 
Actually, the present paper is also devoted to compare in some details 
the two considered models and outline their different microscopic properties. 

Before entering the details of the paper, in the following section, 
for sake of completeness, we briefly explain the definition of 
our lattice gas models and we point out their relations 
to some well known models as the Ising model, 
its Blume-Emery-Griffith counterpart, Edwards-Anderson Spin Glasses 
and driven systems as the Katz-Lebowitz-Spohn model.

\section{Frustrated lattice gas models}

The characteristic properties 
of the Frustrated Lattice Gas we introduced to describe the dynamics 
of dry granular materials (in the regime of high packing densities and 
small shaking amplitude), 
are fully described in Ref.s \cite{NCH,Caglioti}. 

The crucial ingredient of these models is, actually, the presence 
of ``frustration'' in the motion of grains \cite{CH,NCH,Caglioti}. 
The models consist of a system of elongated particles which occupy the 
sites of, say, a square lattice tilted by $45^0$. 
The particles have an internal degree of
freedom $S_i =\pm 1$ corresponding to two possible orientations on 
the square lattice. 
Nearest neighbour sites can be both occupied only if particles do 
not overlap (i.e., they have the right reciprocal orientation) 
otherwise they have to move apart. 
In our models particles undergo a driven 
diffusive Monte Carlo (MC) dynamics. In absence of vibrations they are 
subject only to gravity 
and they can move downwards always fulfilling the non overlap  condition.
The presence of vibration is introduced by also allowing particles to
diffuse upwards with a probability $p_{up}$. The quantity 
$x_0=p_{up}/p_{down}$ (with $p_{down} =1-p_{up}$), as we will see, 
is related to an effective temperature \cite{NC_rev} and plays the role of the 
experimental tap vibrations intensity. 

Our models can be interestingly mapped on standard lattice gas models 
of Statistical Mechanics, whose Hamiltonian has an hard core repulsion term 
($J\rightarrow\infty$): 
$H_{HC}=J\sum_{\langle ij\rangle } f_{ij}(S_i,S_j)n_i n_j$, 
where $n_i=0,1$ are occupancy variables, 
$S_i=\pm 1$ are the above spin variables 
associated to the orientations of the
particles, $J$ represents the infinite repulsion
felt by the particles when they overlap. 
The hard core repulsion function 
$f_{ij}(S_i,S_j)$ is 0 or 1 depending whether the orientations 
$S_i,S_j$ of neighbours is allowed or not. 

The choice of $f_{ij}(S_i,S_j)$ depends on the particular model. 
Here we consider two models: the Tetris and the 
Ising Frustrated Lattice Gas (IFLG). 
In the Tetris model $f_{ij}(S_i,S_j)$ is given by: 
$f^{Tetris}_{ij}(S_i,S_j)=1/2(S_i S_j -\epsilon_{ij} (S_i+ S_j) +1)$
here $\epsilon_{ij}=+ 1 $ for bonds along one direction of the lattice and 
$\epsilon_{ij}=- 1 $ for bonds on the other. 
This corresponds to a generalised Blume-Emery-Griffith Hamiltonian, and  
has an ``antiferromagnetic'' equilibrium phase diagram. 
In order to have a non trivial behaviour 
the dynamics of the Tetris has a crucial {\em purely kinetic constraint}: 
particles can flip their ``spin'' only if many of their own 
neighbours are empty (3, in our simulations), as much as in 
facilitated kinetic Ising models \cite{kob}.

Real granular media may have more disorder due to broader grain 
shape distribution or to absence of a regular underlying lattice, and 
each grain moves in the disordered environment generated by the others. 
To describe this kind of scenario, we previously introduced a model, 
the Ising Frustrated Lattice Gas (IFLG), made of grains 
moving in a lattice with quenched geometric disorder, 
with the following hard core repulsion function:
$f_{ij}^{IFLG}(S_i,S_j)=1/2(\epsilon_{ij} S_i S_j -1)$
where $\epsilon_{ij}=\pm 1$ are quenched random interactions associated to the
edges of the lattice, describing the fact that particles must satisfy the 
geometric constraint of the environment considered as ``practically'' 
quenched. 
The IFLG shows a non trivial dynamics {\em without} 
the necessity to introduce kinetic constraints.
The Hamiltonian of the IFLG exhibits rich behaviours
in connection with those of ``site frustrated percolation''  
\cite{Coniglio} and Spin Glasses \cite{NCH,Arenzon}. 

The other important contribution to the Hamiltonian of a granular media 
we consider, is gravitational energy: 
$H_G=g\sum_i n_iy_i$, 
where g is the gravity constant 
and $y_i$ is the hight of particle $i$
(we set to unity grains mass and lattice spacing). 
The temperature, $T$, of the present Hamiltonian system (with $J=\infty$)
is related to the ratio $x_0=p_{up}/p_{down}$ via the following relation: 
$e^{-2g/T} = x_0$. The adimensional quantity 
$\Gamma\equiv \ln(x_0^{-1/2})= T/g$, seems to  
play the same role as the amplitude of the vibrations 
in real granular matter \cite{NCH}. 

A further term which could appear in the Hamiltonian is a 
coupling between neighbouring 
grains, due, for instance, to the presence of some fluids 
which, due to surface tensions, 
exerts attraction between particles, or electrical charges on grains surfaces 
which might attract or repel others grains \cite{Israelachvili}. 
Thus the full schematic Hamiltonian may have an other term as the following: 
\begin{equation}
H_{SI}=-K_2\sum_{<ij>} n_in_j
\end{equation}
The coupling $K_2$ must be fixed by the specific interaction potential
considered. Thus, the complete Hamiltonian results to be of the form: 
\begin{equation}
H=H_{HC}+H_G+H_{SI}
\label{H_full}
\end{equation}
Below, we will generally consider a system of particles interacting just 
via hard core repulsions, i.e., $K_2=0$, corresponding, for instance, 
to dry non charged granular systems. We will also discuss 
the presence of attraction between neighbouring grains (i.e., $K_2>0$), 
which, we show, does not qualitatively change the general scenario
(at least for $K_2$ not too large with respect to $g\Gamma$). 

It is interesting to notice that the present models are very similar to 
a driven Ising lattice model introduced by Katz, Lebowitz and Spohn 
to describe non equilibrium steady states in fast ionic conductors 
and other systems \cite{KLS,SZ}. One of the relevant differences 
is the boundary conditions imposed on particles motions. 

\begin{figure}[ht]
\vspace{-1.5cm}
\hspace{-1cm}
\centerline{\psfig{figure=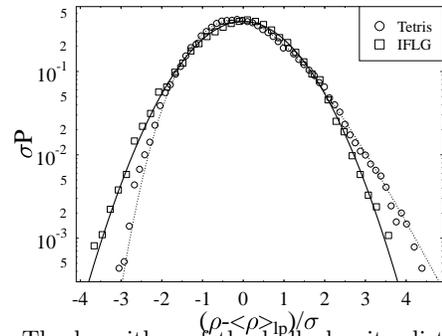,height=7cm,angle=-90}}
\vspace{-1.5cm}
\caption{
The logarithm of the bulk density distribution, $\sigma P(\rho)$, 
recorded after a random pouring of grains in the system box, 
as a function of $(\rho-\lan \rho\ran_{lp})/\sigma$. 
Here $\lan \rho\ran_{lp}$ is the average 
loose packing density and 
$\sigma$ its mean square deviation.
MC data from the IFLG (squares) are well fitted by a Gaussian, 
however, data from the Tetris (circles) clearly show a violation 
of such a behaviour. 
A Gaussian function $(2\pi)^{-1/2}\exp(-x^2/2)$ 
(full line) is shown for comparison. 
We also show an overall Gumbel like fit (dotted line) 
described in the text. 
} 
\label{den_dis_tet}
\end{figure}

\section{Bulk density distribution} 

Monte Carlo simulations of the present models have periodic
boundary conditions along the horizontal direction 
and rigid walls at bottom and 
top (here we consider a two dimensional sample, but analogous results 
are found in three dimensions as shown in Ref.\cite{NCH}). 
After fixing the $\epsilon_{ij}$ (which are random in the IFLG), 
the initial particle configuration is
prepared by randomly inserting particles of given spin 
into the box from its top and then letting them fall down ($p_{up}=0$) 
until the box is filled. 
The two basic Monte Carlo moves (the spin flip and particle hopping) 
are done in random order. 

An important quantity to characterise grains packing 
after such a random insertion, is the bulk density distribution, $P(\rho)$. 
Such a quantity is shown in Fig.\ref{den_dis_tet} for the IFLG and the 
Tetris. The data for IFLG concern a lattice box of size 
$100\times 100$ and for Tetris a box $120\times 1000$ 
and are averaged over 50000 configurations 
(the IFLG size is smaller since we have also 
to average over the $\epsilon_{ij}$). 

The density where $P(\rho)$ has its maximum practically corresponds 
to the average loose packing density of the system $\lan \rho\ran_{lp}$ 
($\lan \rho\ran_{lp}\simeq 0.739$ in the IFLG and 
$\lan \rho\ran_{lp}\simeq 0.751$ in the Tetris \cite{nota_1}), whose 
precise location depends on the model and its linear sizes 
(for instance, boundary effects can change it). 
In Fig.\ref{den_dis_tet} the logarithm of $\sigma P(\rho)$ is plotted 
as a function of $(\rho-\lan \rho\ran_{lp})/\sigma$ for both data sets 
($\sigma$ is the density mean square deviation, 
$\sigma\simeq 0.002$ in the IFLG and $\sigma\simeq 0.004$ in the Tetris).
This way to plot $P(\rho)$ allows a direct comparison between the two 
different distributions. Furthermore, it easily outlines deviations 
from a Gaussian behaviour. Actually, after the above data rescaling, 
a Gaussian variable must have a 
normalised univariate zero mean Gauss distribution function 
(with no adjustable parameters): 
\begin{equation}
P(x)=\frac{\exp(-x^2/2)}{(2\pi)^{1/2}}
\end{equation}
This function is the full line in Fig.\ref{den_dis_tet}. 
Apparently, the data for the IFLG model (squares) seem to reasonably follow 
such a Gaussian shape, a fact which well compares with experimental data 
from Ref.\cite{Novak}. 

The Tetris case, is more controversial from our simulations. 
Data from a $100\times 100$ sized system seem to follow 
the Gaussian distribution found for the IFLG. They are not plotted in 
Fig.\ref{den_dis_tet} for sake of clarity. 
In Fig.\ref{den_dis_tet} are shown, instead, Tetris data (circles) 
for the above cited 
largest system size (about $10^5$ grains) we could simulate 
(their statistics is even better than the one for the smaller system). 
Interestingly, in this case $P(\rho)$ is {\em not} Gaussian. 
For density values below $\lan \rho\ran_{lp}$, $P(\rho)$ rapidly falls off 
while above $\lan \rho\ran_{lp}$ it surprisingly shows an almost exponential 
tail (see Fig.\ref{den_dis_tet}). 
In Fig.\ref{den_dis_tet}, we also plot the overall fit 
obtained by using a Gumbel-like function:
\begin{equation}
P(x)=P_0\{\exp[-a(b+x)-e^{-a(b+x)}]\}^m
\end{equation}
where $P_0\simeq 37$, $a\simeq 0.5$, $b\simeq 0.2$ and $m\simeq 4.5$. 
The above Gumbel-like distribution is interesting for its  
relations to extreme values statistics \cite{Gumbel}, but 
it is difficult to state whether such an approximate three parameters fit 
($P_0$ is fixed by the normalisation) plays an important role 
in the present context. It should be said, however, that 
experimentally a non Gaussian $P(\rho)$ is usually found after 
long ``tapping'' of the system \cite{Novak}. 
A finite amount of CPU time has not allowed us to go that far. 
Further theoretical and experimental 
work must be devoted to understand the present results. 

\begin{figure}[ht]
\centerline{\psfig{figure=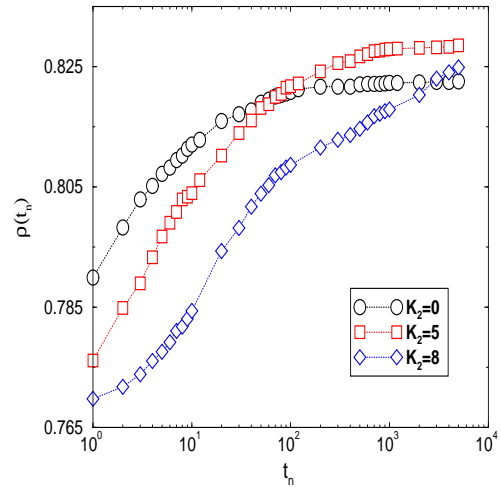,height=6.5cm,width=6.5cm,angle=-90}}
\caption{
The presence of attractive interactions among grains, may be simulated 
by turning on $K_2$ in our full Hamiltonian. 
We show here how such an attraction affects compaction. 
We plot the bulk density of the Tetris, $\rho$, as a function of $t_n$, 
the tap number, for different values of $K_2$ (given in units of $g\Gamma$). 
The system size is $30\times 60$ and $e^{-2/\Gamma}=x_0=0.05$
(similar results are found for $x_0\in[0.01,0.1]$). 
Data are averaged over 240 taps sequences. 
The grains attraction makes the initial relaxation slower, 
also if the long times value of density is higher the higher is $K_2$. 
} 
\label{humid}
\end{figure}

\section{Compaction in presence of grains attractive interactions}

After having described the properties of the starting random loose packed 
configuration (for other details see \cite{NCH}), 
in this Section we start discussing the phenomenon of 
compaction. This plays an important role in understanding grains dynamics 
since it clearly shows the basic mechanisms underlying 
dynamical processes in gently shaken granular media subject to gravity
\cite{Knight}. 

Referring to recent experiments on compaction dynamics, 
here we study the phenomenon of density increase due to tapping. 
In real experiments a ``tap" is the shaking of a container filled with grains 
by pulsed vibrations of given duration and amplitude \cite{Knight}.
In our MC simulations, in each single tap 
we apply vibrations of a given amplitude $x_0$ 
to particles for a given, short, duration $\tau$ 
(i.e., for $t\in[0,\tau]$ we fix $x(t)=x_0=const.$) 
\cite{nota_t}. 
Then the system is let to find a stationary state for a time $t_{repose}$ 
in which $x(t)=0$ ($t_{repose}$ is chosen to be much longer than any 
relaxation times in absence of shaking \cite{NCH}). 
After each tap we measure the static bulk density of the system 
$\rho(t_n)$ ($t_n$ is the $n$-th tap number). 
For this Monte Carlo experiment with the IFLG model 
we considered a system of size $30 \times 60$, averaged over $32$ 
different $\epsilon_{ij}$ configurations, and fix $\tau=32$. 
We refer to Ref.\cite{NCH} for further details on the MC dynamics. 

To describe experimental observations 
about grain density relaxation under a sequence of taps  
a logarithmic law was proposed in Ref.~\cite{Knight}:
$\rho(t_n)=\rho_{\infty}-\Delta\rho_{\infty}/[1+B ~ \ln(t_n/\tau_1+1)]$. 
This law has proved to be satisfied very well by relaxation data in
the present IFLG model as shown in Ref.\cite{NCH}, 
which can be excellently rescaled with experimental data.

The results from IFLG and Tetris are surprisingly similar \cite{NCH,Caglioti}, 
but the asymptotic density $\rho_{\infty}$ in the Tetris is numerically 
indistinguishable from 1, thus almost independent on $x_0$, a fact in 
contrast with both IFLG and experimental results from Knight et al. 
\cite{Knight}. 

In order to have a full comparison of our models 
with a variety of experimental results and to understand the effects of 
small perturbations on the grain interaction potential on the general 
compaction scenario, we also dealt with 
the compaction of systems whose neighbouring grains feel a finite 
attraction. This is the case, for instance, in presence of grains 
with interstitial fluids, which may generally affect the system properties 
\cite{Israelachvili,castel}. Thus, we turn on 
$K_2$ in our Hamiltonian eq.(\ref{H_full}). 

The results of density compaction of interacting attractive grains are 
shown in Fig.\ref{humid}. 
The presence of a finite $K_2$ doesn't alter the general logarithmically 
slow features of the dynamics in the considered range $K_2/g\Gamma\in[0,8]$, 
but interesting new phenomena appear. 
Fig.\ref{humid} concerns the case of the Tetris: the figure 
shows, for reference, the density compaction, $\rho(t_n)$, 
of a system without inter-grains attraction (circles) 
during a sequence of MC taps as those described above; 
it shows moreover the compaction of other two systems 
with finite inter-grains attractions, $K_2$: the interaction strengths are 
expressed as $K_2=5$ (squares) and $K_2=8$ (triangles) in units of  $g\Gamma$, 
which is the relevant physical energy scale. 
One can see that two seemingly opposite behaviours are present: 
the grains attraction makes the relaxation much slower 
also if the long times value of density is higher the higher is $K_2$. 
For $K_2=8g\Gamma$, some overall qualitative change in the density relaxation 
curve is also appreciable. 

\subsection{Evolution by ``jumps'' and ``persistence times''}

The density evolution curves shown in Fig.\ref{humid} are obtained as an 
average over many taps sequences and for such a reason they look regular. 
However, the shape of the curve $\rho(t_n)$ in each 
single run is interesting in itself since it is highly irregular 
and its evolution determined by sudden changes. 
Actually, during several consecutive ``taps'' the systems seems frozen 
in the same state, 
then, after a new tap, abruptly its bulk density jumps to a new value (which 
only on average is higher that the previous). 
During such seemingly ``dead'' time intervals the system persists in the 
same state, with a given bulk density. 
We call ``persistence times'', $t_p$, the duration of these intervals.  

\begin{figure}[ht]
\vspace{-1.5cm}
\hspace{-1cm}
\centerline{\psfig{figure=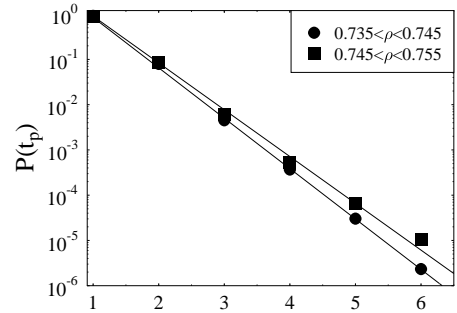,height=7cm,angle=-90}}
\vspace{-1.5cm}
\caption{
The logarithm of the ``persistence time'' distribution, $P(t_p)$,
as a function of $t_p$. $t_p$  
is the number of elapsed taps between two consecutive changes in the 
bulk density of the IFLG during sequences of taps (here with $x_0=0.0001$). 
The lower data (circles), which are recorded when the system density is 
in the interval $[0.735,0.745]$, and the upper data (squares), 
recorded in the interval $[0.745,0.755]$, 
show an exponential behaviour (full lines). 
} 
\label{pers}
\end{figure}

Persistence times have an interesting distribution, $P(t_p)$, 
which depends on the system density $\rho$. In order to 
increase the accuracy, data plotted in Fig.\ref{pers} are values 
obtained after averaging in a given density interval. 
At low densities (close to $\lan\rho\ran_{lp}$), 
$P(t_p)$ certainly has exponential shape: 
\begin{equation}
P(t_p)\sim e^{-t_p/t_p^0}
\end{equation}
In the interval, $0.735<\rho <0.745$, we find $t_p^0\simeq 0.40$. 
In a higher densities interval $0.745<\rho <0.755$, $P(t_p)$ 
seems to becomes broader (see Fig.\ref{pers}). 
It is extremely difficult to collect sufficiently clean data 
for high values of $t_p$ in the high density region 
(due to the required very long CPU times) in order to investigate 
such a ``broadening'' of the $P(t_p)$. An exponential fit (with 
$t_p^0\simeq 0.42$) is still possible \cite{nota_Ptp}. 
These results are plotted in Fig.\ref{pers}, where we show data for 
the IFLG (size $30\times 60$) 
recorded during sequences of taps with $x_0=0.0001$ ($K_2=0$). 
Similar behaviours of $P(t_p)$ are observed in the region 
$x_0\in[10^{-4},10^{-1}]$, showing that not only the overall 
properties of the dynamics do not change in such a ``shaking'' 
interval. 

We discuss below, from a microscopical point of view, how 
density changes are made possible by the local reorganisation 
of the structure of grains clusters. 

\begin{figure}[ht]
\centerline{\psfig{figure=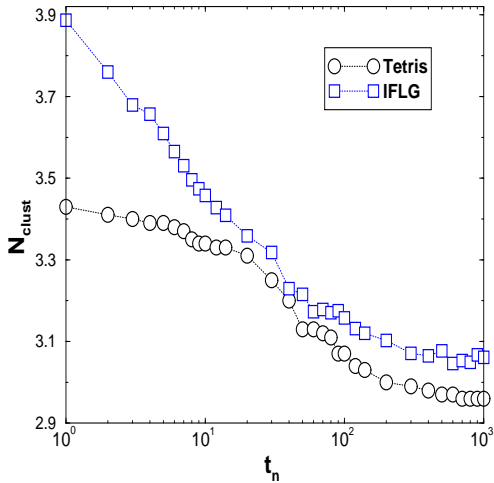,height=6.5cm,width=6.5cm,angle=-90}}
\caption{
In our lattice systems, we define a ``cluster" as a set of 
nearest neighbour grains. 
In the present figure (concerning a lattice of size $30\times 60$), 
we show the number of clusters, $N_{clust}$,  as a function of 
the taps index $t_n$ during a MC sequence of taps in 
both Tetris (filled circles) and IFLG (empty squares). 
In both kind of models are there very few clusters for unit length 
and their number very slowly decreases with $t_n$.
The data are averaged over 100 taps sequences with $x_0=0.1$.
} 
\label{n_clus}
\end{figure}

\section{Domains of grains in the pack}

Exploiting the microscopic character of our models, we have access 
to the details of the above compaction process and we explore the 
grains packing structure. 

In our lattice systems we can define a ``cluster" of grains 
as a set of nearest neighbouring particles. Due to grains shapes 
incompatibilities, empty sites can be present in the lattice pack. 
For instance in the Tetris, clusters can be pictorially described to be 
mainly thick ``vertical'' clumps separated by almost linear 
sequences of holes. 
Actually, the above definition of ``clusters'' must be refined to be 
applicable to a real granular medium, but on a lattice it is 
the most natural. 

\begin{figure}[ht]
\centerline{\psfig{figure=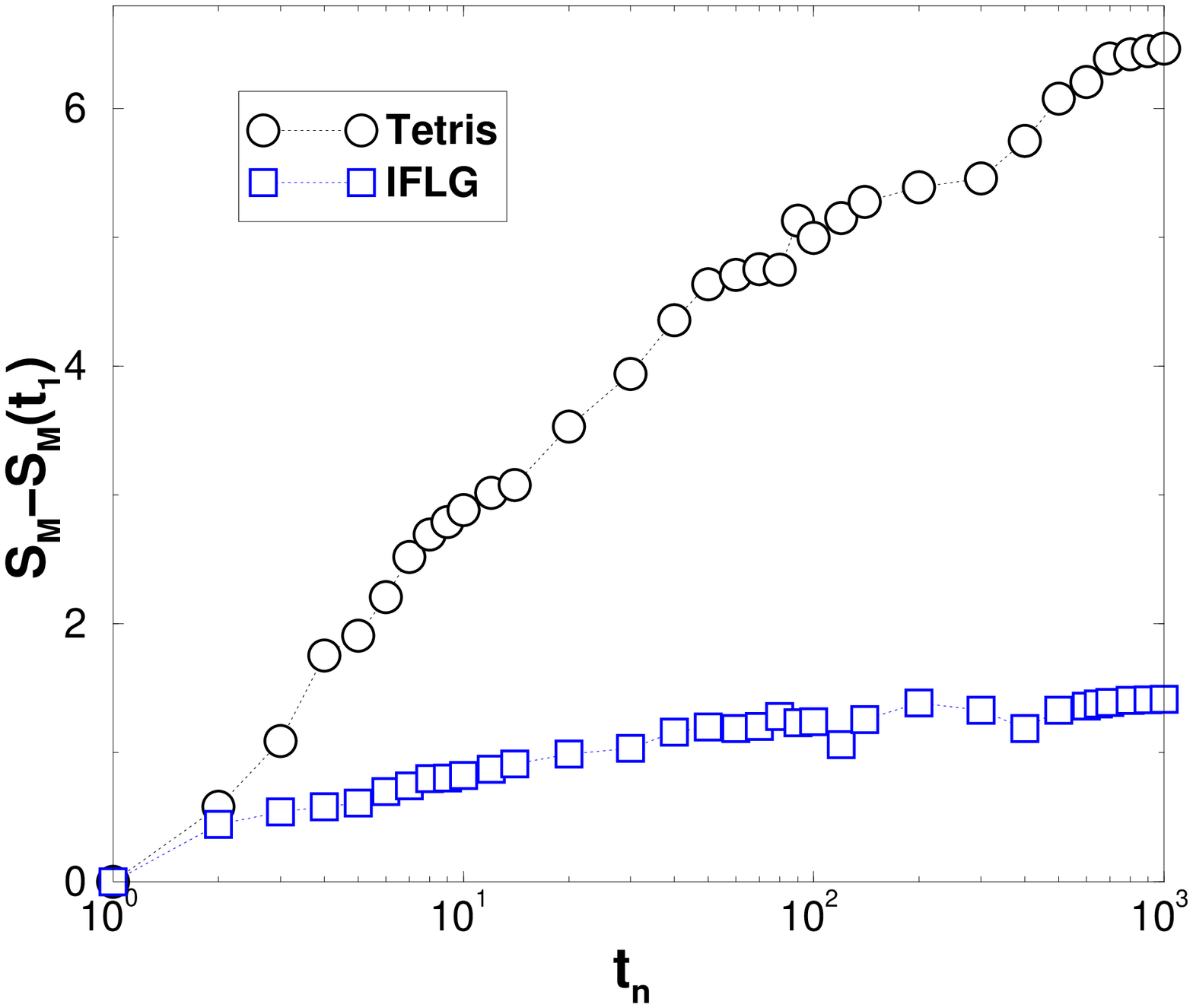,height=6.5cm,width=6.5cm,angle=-90}}
\centerline{\psfig{figure=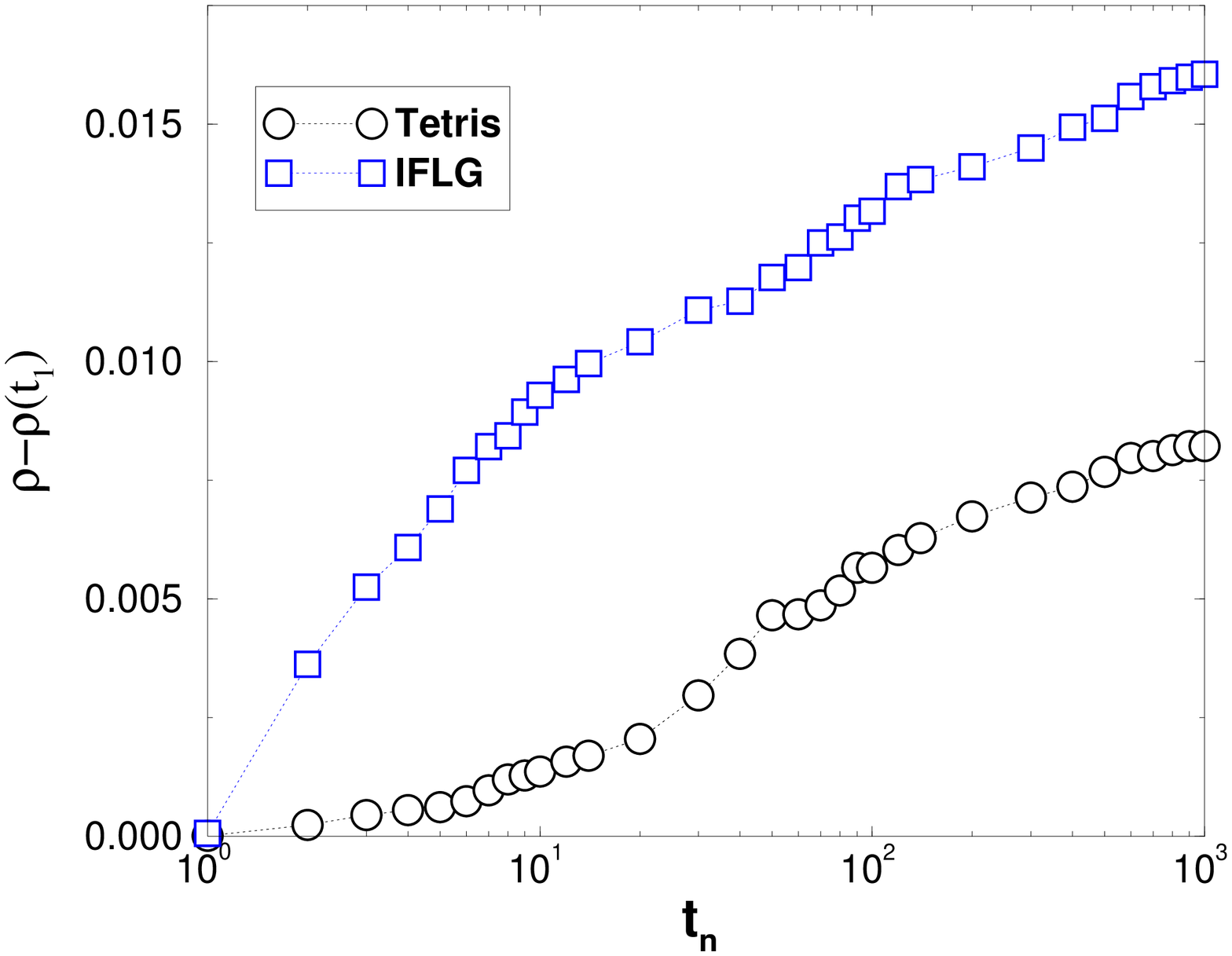,height=6.5cm,width=6.5cm,angle=-90}}
\caption{
{\em Top:} we plot the difference of 
the average sizes of the largest cluster in the system, $S_{M}$, 
minus the same quantity after the first tap, $S_{M}(t_1)$, 
as a function of the taps index $t_n$, during a MC taps sequence 
with $x_0=0.1$ in both Tetris (circles) and IFLG (squares). 
The box size is $30\times 60$. 
The values at $t_1$ are: $S_{M}(t_1)=919$ in the IFLG, and 
$S_{M}(t_1)=687$ in the Tetris. 
{\em Bottom:} for a clear comparison, we also plot the bulk density variation 
of the same systems during the same taps sequences. 
The values at $t_1$ are: $\rho(t_1)=0.743$ in the IFLG, and 
$\rho(t_1)=0.830$ in the Tetris. 
Data are averaged over 100 taps sequences. \\ 
Also if the overall behaviour is similar, 
the compaction process of the two considered models is 
microscopically very different. In the Tetris it is originated by 
the growth of the largest cluster at expense of the smaller ones, 
as in spinodal decomposition. In the IFLG on the contrary, the 
growth of the largest cluster is extremely weak: the compaction is 
mainly due to the ``expulsion'' of holes from the largest clusters 
and the consequent optimisation of grains arrangement.
} 
\label{smax}
\end{figure}

In the randomly prepared initial state are there typically very few 
clusters per unit length (in a $30\times 60$ 
sized lattice one has a total of 3-4 clusters, see Fig.\ref{n_clus}). 
By tapping, the number of clusters, $N_{clust}$, approximately logarithmically 
decreases with the taps index $t_n$ (see Fig.\ref{n_clus}). 
In the same time the larger cluster grows and the others shrink 
(see Fig.\ref{smax}). The data shown in Fig.\ref{n_clus} and Fig.\ref{smax} 
refer to tapping sequences with ``amplitude'' $x_0=0.1$, and analogous 
results were found for $x_0=0.01$.

The analysis of cluster properties reveals interesting facts about the 
compaction mechanisms and outlines a main difference 
between the compaction in IFLG and Tetris: 
it concerns the relative size of the few present clusters. 
In the IFLG the second largest cluster is, since the 
beginning, made of very few grains ($O(1)$); in the Tetris, the first 
and second largest clusters are instead, 
at the beginning, almost of the same order of magnitude. 

During the taps sequence, in the Tetris the largest cluster 
grows at expense of the smaller ones. As shown in Fig.\ref{smax}, after 
$10^3$ taps, a $1\%$ increase of the bulk density, $\rho(t_n)$, 
is in direct  correspondence 
with a $1\%$ increase in the largest cluster mass, $S_M(t_n)$. 
This mechanism strictly recalls 
ordered domains growth in spinodal decomposition \cite{Bray}. 
As a matter of fact, the Tetris is a ``kinetically constrained'' 
antiferromagnet and our clusters  explicitly correspond 
to the definition of the ``Fisher droplets'' in the system 
\cite{droplets}. In the IFLG, on the contrary the 
growth of the largest cluster is extremely weak
(see Fig.\ref{smax}) since the other few clusters size is 
already of the order of very few grains: now 
a $2\%$ increase in the bulk density corresponds to only a $0.1\%$ increase 
in the largest cluster mass. In the latter case the 
compaction of the system is, thus, 
mainly due to the ``expulsion'' of holes from the largest cluster 
and the consequent optimisation of grains spatial arrangement. 

In the present perspective the basic mechanisms of domains growth under 
gravity in the IFLG and Tetris seem to be very different. 
To understand what scenario is closer to reality, if any of the present, 
an experimental investigation of this kind of properties would be 
very important. 

The description of compaction with a diffusion equation with a density 
dependent non linear diffusion coefficient was proposed in 
Ref.s~\cite{Ben-Naim,Caglioti}. It resembles analogous behaviours 
found close to the glass transition in glass formers \cite{Corberi}. 

\section{Conclusions}

Summarising, the central body of the present paper has dealt with 
the dynamical behaviours of two frustrated lattice gas models 
(the IFLG and Tetris) introduced to describe gently vibrated granular media 
\cite{NCH,Caglioti}. These microscopic models are characterised by 
a gravity driven 
diffusive dynamics where the basic ingredient is the presence of geometric 
frustration in particles rearrangement. 
Interestingly, although granular media are non-thermal dissipative systems, 
the present models can be casted in the 
Hamiltonian formalism of standard lattice gases of Statistical Mechanics 
in presence of an effective temperature (different from zero during shaking). 

To characterise the 
grains packing structure we have recorded the loose packing 
density distribution function, $P(\rho)$, which in the case of the 
IFLG well compares with experimental results. More interesting, but 
also more controversial is the non Gaussian $P(\rho)$ recorded in the Tetris.

Also in presence of attractive surface grains interactions,
the models exhibit a slow compaction when subject to gentle shaking in 
presence of gravity, a compaction extremely close to what is experimentally 
observed in granular packs \cite{Knight}. 
During such a process, the dynamics becomes slower and slower since 
grains self-diffusivity decreases with density \cite{NCH}. At high densities 
the system approaches a structural arrest at a ``jamming" transition point, 
where self-diffusivity becomes zero, a fact very similar to the ``freezing 
transition" in glass formers \cite{NC_aging,N_fdt,NC_rev}. 
Interestingly in the IFLG, the structural arrest due to self-diffusivity 
suppression coincides with a spin glass transition \cite{NCH}. 

The microscopic nature of the models allows the study of the properties 
of their internal packing structure, as grains clusters, which give 
a deeper understanding of the basic mechanisms underlying compaction. 
The study of ``grains clusters" shows that 
in the IFLG model compaction originates from the reduction of the number of 
holes in the only present large cluster 
by optimisation of grains spatial arrangement; 
in the Tetris, instead, compaction  
stems from the growth of the largest cluster mass itself 
at expense of the smaller ones, as in spinodal decomposition. 
Actually, this result points towards the necessity of a better understanding 
of the real long time dependence of compaction in the Tetris and IFLG. 

Many of the results presented here are still waiting for an important 
experimental verification. 

\smallskip

Work partially supported by the TMR Network Contract ERBFMRXCT980183 and 
MURST-PRIN 97.


\begin{thebibliography}{70}
\bibitem{nota_1} Notice that the densities in the IFLG model are usually 
lower than in the Tetris.
\bibitem{nota_Ptp} Notice that 
a power law interpolation of the tail is approximately consistent  
with the same accuracy of an exponential fit. 
Using a function $P(t_p)\sim t_p^{-\gamma'}$, 
for $P(t_p)$ in the interval $0.745<\rho <0.755$, 
from a fit over just half a decade, we get, for instance, 
an exponent (very large) $\gamma'\sim 9$.
\bibitem{nota_t} Time is measured in such a way that unity corresponds 
to an average update of every degree of freedom in the system.
\end{thebibliography}
\end{document}